# A Survey of Digital Privacy Rights Under CISA

Matthew Lemanski
College of Computer and Information Science
Northeastern University, Boston, MA
lemanski@ccs.neu.edu

*Abstract* – The recent passing of the Cybersecurity Information Sharing Act of 2015 introduces a new framework for information sharing between private and US government entities with the expressed intent to identify cybersecurity threats. This is the latest in a series of similar bills that have been introduced to Congress over the last several years. While each of the previous standalone bills were defeated following widespread public resistance, the latest edition was included as an amendment to the United States' 2016 spending bill. This means that any dissenting congressmen unwilling to pass the spending bill with the CISA rider would be willing to risk another government shutdown due to the inability to come to terms on the budget measures. This paper seeks to explore the potential impacts of the measures introduced or enabled by CISA, and consider the formalization of digital privacy rights in an increasingly online and monitored world.

## I. INTRODUCTION

The Cybersecurity Information Sharing Act of 2015 (CISA) [1] was enacted on December 18, 2015 as an amendment to the US government's 2016 omnibus spending bill [2, 3]. CISA introduces a new framework enabling automated dissemination of perceived cybersecurity threats between private and federal entities. Similar standalone bills (Stop Online Piracy Act [4], Cyber Intelligence Sharing and Protection Act [5], and CISA [1]) were proposed to congress in recent years with mixed success; each was eventually stricken down by opposition to the wide scope, vague language, and recent revelations of systemic government surveillance [6]. By including the latest iteration of CISA as a rider to the omnibus spending bill, any congressman compelled to vote against the information sharing act would need to vote against the spending bill as a whole. Failure to pass the spending bill would risk a government shutdown like the one experienced in 2013 [7], a measure some congressmen opposing CISA may not have been willing to make.

CISA defines a framework for sharing information about perceived cybersecurity threats between internet companies and government entities [8]. Now that CISA is officially a law in the United States, what developments can we expect to see in internet-based service providers? What sort of information will internet companies be sharing with the government? Who will have access to this information? The paper seeks to answer those questions and explore the use of personally identifiable information (PII) in the system enabled by CISA.

## II. CISA'S PROVISIONS

The information sharing framework outlined by CISA enables government and private entities to share data with each other with the expressed intent of identifying and subverting cybersecurity threats [2, 3]. The act does not require companies to share information, but rather puts in place a system to make sharing possible.

Under this act companies monitoring their traffic would, upon identifying a potential cyber threat, automatically forward the traffic information to a system managed by the Department of Homeland Security (DHS), which would then in turn share the information directly with other government entities [9], such as the National Security Agency (NSA) or Central Intelligence Agency (CIA). In previous iterations of CISA the DHS was required to vet the received traffic by scrubbing any PII, but it remains unclear what entity would be responsible for ensuring that PII is removed from the collected data [10]. Furthermore since the sort of PII will vary depending on the traffic, it is unclear to what extent PII will be removed. CISA states that shared information shall not identify an individual, and that only traffic directly related to a cybersecurity threat will be shared, but how either of those thresholds is determined is undefined.

CISA also allows companies to share data without the risk of liability [2]. This could remove some companies' obligation to protect their users, which could conceivably lead users to find alternative service providers.

## III. RISK OF INFORMATION SHARING

As with previously proposed online traffic sharing laws (i.e. SOPA, CISPA), CISA has generated controversy from internet companies and privacy advocates alike [11-13]. Even if the groups implementing CISA manage an infallible implementation, the privacy advocates contend that an internet user has the right to privacy, and that the government has no right to collect and study a user's traffic [14]. The picture is further muddied by the question of who owns the content a user uploads any social media or other website [15].

Furthermore, revelations in recent years have revealed that the US has been involved with several schemes that attempt to subvert a user's online privacy. From bulk collection of phone records [16], to bulk collection of internet traffic [17, 18], to attacking common online encryption methods [19, 20], the US has proven that it is interested in analyzing as much internet traffic as possible. Even if its goal is truthfully an attempt to "protect… an information system from a cybersecurity threat", as stated in the CISA text, its previous attempts to undermine encryption standards leave it in an untrustworthy position to manage bulk traffic data.

## IV. RIGHT TO DIGITAL PRIVACY

The revelations of the US government's pervasive surveillance raise questions about how several of a US citizen's constitutional rights, including the First, Fourth, Fifth, and Fourteenth Amendments [21-24], are protected in a world with so much personal information available online [25]. The application of the Fourth Amendment, which protects a citizen's rights against unreasonable searches and seizures, appears to be the most relevant with the enactment of CISA.

The concept of the Fourth Amendment requires that a law enforcement agency gathers enough evidence against a person to obtain a warrant before ever attempting to search or seize a person's property [22]. Without probably cause, the agency should not be able to obtain permission to search the person's property. But how does this apply to digital communications that are stored by a phone company or an internet service provider? While earlier Supreme Court decisions established a person's right to privacy while engaging in a telephone conversation [26], the Electronic Communications Privacy Act of 1986 (ECPA) and the PATRIOT Act of 2001 together allow federal agencies to search telephone, e-mail, financial, and other records without requiring a warrant [27, 28]. Plenty have argued that these laws are unconstitutional and a violation of Fourth Amendment rights [29]; in fact the US government allowed the provisions enabling bulk collection of phone records to expire just this year [30]. Considering the ECPA's and PATRIOT Act's questionable usage, the recent revelations of the NSA's subversive actions, and the today's increasingly mobile and internet-enabled technologies, several groups have begun advocating for new digital privacy rights to protect the growing amount of user data [31].

One proposal is the introduction of new legislation to modify the ECPA in order to guarantee a user's right to protect their digital information under the Fourth Amendment [31]. Such a law would ensure that user information, including PII, would be protected from all analysis unless the agency is able to show probably cause and obtain a warrant. Protecting digital user information under the Fourth Amendment would certainly go against the system put forth in the CISA act, meaning the recently enacted legislation would need to be overhauled, if not repealed, to conform to the proposed digital privacy standards. Modifications to the CISA legislation could be enacted to share anonymous traffic patterns could still be leveraged to identify cybersecurity threats. Threat patterns revealed through traffic such as bot net activities would still be readily identifiable from the aggregate picture, and without violating sensitive user data.

## V. CONCLUSION

The recently passed CISA law, included in the US government's 2016 omnibus spending bill, enables a framework for sharing information between private companies and US government entities. Questions remain about how CISA will work in practice, since it is unclear how PII will be identified and removed, and by whom. Though CISA's advertised intent is to identify and stop cybersecurity threats, the government's history shows that it has an interest in maintaining an ability to access as much data transiting the internet as it can. The current laws regarding analysis of stored electronic communications appear to already be quite favorable to the government, and arguably violating the Fourth Amendment. By enacting legislation to cover a person's electronic information under the Fourth Amendment the United States could conceivably guarantee that person's right to digital privacy while maintaining the ability to detect cybersecurity threats.